\documentclass[aps,twocolumn,showpacs,superscriptaddress,amsmath]{revtex4}
\usepackage{graphics,epsf,colordvi}
\newcommand{\bea}{\begin{eqnarray}}
\newcommand{\beq}{\begin{equation}}
\newcommand{\eea}{\end{eqnarray}}
\newcommand{\eeq}{\end{equation}}
\begin{document}
\title{Entanglement as an indicator of a geometrical crossover
in a two-electron quantum dot in a magnetic field}
\author{R. G. Nazmitdinov}
\affiliation{Departament de F{\'\i}sica,
Universitat de les Illes Balears, E-07122 Palma de Mallorca, Spain}
\affiliation{Bogoliubov Laboratory of Theoretical Physics,
Joint Institute for Nuclear Research, 141980 Dubna, Russia}
\author{N. S. Simonovi\'c}
\affiliation{Institute of Physics, University of Belgrade, P.O.
Box 57, 11001 Belgrade, Serbia}
\begin{abstract}
We found that  a downwardly concave entanglement evolution of the ground state
of a two-electron axially symmetric quantum dot testifies that a shape
transition from a lateral to a vertical localization of two
electrons under a perpendicular magnetic field takes place.
Although affected, the two-electron
probability density does not exhibit any prominent change.
\end{abstract}
\pacs{03.67.Bg, 73.21.La, 73.22.-f}
\date{\today}
\maketitle

Quantum phase transitions (QPTs) in many-body systems, driven by
quantum fluctuations at zero temperature, attract a considerable
attention  in recent years \cite{Sad}. They are recognized  as
abrupt changes of the ground state of  many-body system with
varying a non-thermal control parameter (magnetic field, pressure
etc) in the Hamiltonian of the system. Although the phases should
be characterized by different types of quantum correlations on
either side of a quantum critical point (actual transition point),
in some cases they cannot be distinguished by any local order
parameter. Particular examples are the integer and fractional
quantum Hall liquids \cite{Hall} which cannot be understood in
terms of the traditional description of phases based on symmetry
breaking and local order parameters. Nowadays, there is a growing
interest in using quantum entanglement measures for study such
transitions \cite{am} and, in general, quantum correlations in
many-body systems \cite{TMB2011}.

According to a general wisdom, various phases could exist in
mesoscopic systems such as quantum dots (QDs) at different
strengths of the applied magnetic field.
Indeed, if an axially-symmetric QD is placed under a perpendicular
magnetic field, one observes the orbital momentum and spin
oscillations of the ground state of a QD by increasing the field
strength \cite{RM}. At certain field range the oscillations disappear and it
is believed that the electrons form a finite-size analogue of
infinite integer quantum Hall liquid. This fully polarized state
in a QD is called the maximum density droplet. It is widely
accepted that a further increase of the magnetic field should lead
to the formation of the Wigner molecule, a finite-size analogue of
the Wigner crystallization of the homogeneous electron gas. While
the rotational symmetry of the ground state is expected to be
preserved at the transition between these phases, specific excited
(rotational) states should appear after the transition point
\cite{RM,mak,yan}. These states are not observed yet, and one
needs further to develope  experimental as well as theoretical
tools to detect such transitions which may be associated with
QPTs. Although finite systems can only show precursors of the QPT
behaviour, they are also important for the development of the
concept.

Two-electron QDs being realistic non-trivial systems
are, in particular, attractive. In contrast to 
many-electron QDs, 
their eigenstates can be obtained very accurately, 
or in some cases exactly (cf \cite{kais,r1}). In fact,
two-electron QDs become a testing ground for various approaches
used for description of many-electron QDs. We will show that a
quantum entanglement enables us to detect a geometrical (shape)
transition in the ground state of two interacting electrons
confined in a three-dimensional (3D) quantum dot under a magnetic
field. It turns out that such a transition might be
associated with a quantum phase transition.

We carry our analysis by means of the numerical diagonalization
of the Hamiltonian
\beq \label{ham} H = \sum_{j=1}^2 \bigg[ \frac{1}{2m^*\!}\,
\Big({\bf p}_j \!-\! \frac{e}{c} {\mathbf A}_j \Big)^{\! 2} +
U({\mathbf r}_j) \bigg] + \frac{k}{\vert\mathbf{r}_1 \!-\!
\mathbf{r}_2\vert}+ H_\mathrm{spin}.
\eeq
Here $k = e^2/4\pi\varepsilon_0\varepsilon_r$ and $H_{\it spin} =
g^*\mu_B(\mathbf{s}_1+\mathbf{s}_2)\!\cdot\!\mathbf{B}$ describes
the Zeeman term, where $\mu_B = |e|\hbar/2m_ec$ is the Bohr
magneton. As an example, we will use the effective mass
$m^*=0.067m_e$, the relative dielectric constant $\varepsilon_r =
12$ and the effective Land\'e factor $g^*=-0.44$ (bulk GaAs
values). For the perpendicular magnetic field we choose the vector
potential with gauge ${\mathbf A} = \frac{1}{2} \mathbf{B} \times
\mathbf{r} = \frac{1}{2}B(-y, x,0)$. The confining potential is
approximated by a 3D axially-symmetric harmonic oscillator
$U(\mathbf{r}) = m^* [\omega_0^2\,(x^2 \!+ y^2) +
\omega_z^2z^2]/2$, where $\hbar\omega_z$ and $\hbar\omega_0$ are
the energy scales of confinement in the $z$-direction and in the
$xy$-plane, respectively.

Introducing the center of mass (CM) and relative coordinates:
$\mathbf{R} = \frac{1}{2}(\mathbf{r}_1 + \mathbf{r}_2)$ and
$\mathbf{r}_{12} = \mathbf{r}_1 - \mathbf{r}_2$, -- one  separates
the Hamiltonian~(\ref{ham}) into the CM and relative motion terms $H =
H_{\rm CM} + H_{\rm rel}$ (the Kohn theorem \cite{kohn}). The CM term
is described by the oscillator Hamiltonian with the mass ${\cal M}
= 2m^*$ and frequencies  of the one-particle confining potential
$U$. The Hamiltonian for relative motion in cylindrical
coordinates takes the form
\begin{equation}
H_{\rm rel} = \frac{1}{2\mu}\Big(p_{\rho_{12}}^2 \!+
\frac{\ell_z^2}{\rho_{12}^2} + p_{z_{12}}^2\!\Big) +
\frac{\mu}{2}(\Omega^2 \rho_{12}^2 + \omega_z^2 z_{12}^2) +
\frac{k}{r_{12}} - \omega_L \ell_z, \label{relham}
\end{equation}
where $\mu = m^*/2$ is the reduced mass, $\ell_z$ ($\to
-i\hbar\partial/\partial\varphi$) is the projection of angular
momentum for relative motion and $\rho_{12} = (x_{12}^2 +
y_{12}^2)^{1/2}$, $\varphi = \arctan(y_{12}/x_{12})$, $r_{12} =
(\rho_{12}^2 + z_{12}^2)^{1/2}$. Here, $\omega_L = |e|B/2m^*\!c$
is the Larmor frequency, and the effective lateral confinement
frequency $\Omega = (\omega_{L}^{2} + \omega_{0}^{2})^{1/2}$
depends through $\omega_{L}$ on $B$.

The total two-electron wave function
$\Psi(\mathbf{r}_1,\mathbf{r}_2) = \psi(\mathbf{r}_1,\mathbf{r}_2)
\chi(\sigma_1,\sigma_2)$ is a product of the orbital
$\psi(\mathbf{r}_1,\mathbf{r}_2)$ and spin
$\chi(\sigma_1,\sigma_2)$ wave functions.
Due to the Kohn theorem, the orbital wave function is factorized
as a product of the CM and the relative motion wave functions
$\psi(\mathbf{r}_1,\mathbf{r}_2) =
\psi_\mathrm{CM}(\mathbf{R})\,\psi_\mathrm{rel}(\mathbf{r}_{12})$.
The parity of $\psi_\mathrm{rel} (\mathbf{r}_{12})$ is a good
quantum number as well as the magnetic quantum number $m$, since
$\ell_z$ is the integral of motion.

The CM eigenfunction is a product of the Fock-Darwin state
(the eigenstate of a single electron in an isotropic
2D harmonic oscillator potential in a perpendicular
magnetic field) \cite{Fock} in the $(X,Y)$-plane
and the oscillator function in the $Z$-direction
(both sets for a particle of mass ${\cal M}$). In this paper we
consider the lowest CM eigenstate with the projection of CM
angular momentum equal zero.

Since the Coulomb interaction
($k \neq 0$) couples the motions in $\rho_{12}$ and
$z_{12}$-directions,
the eigenfunctions of the Hamiltonian
for relative motion (\ref{relham}) are expanded in the basis of
the Fock-Darwin states $\Phi_{n,m}(\rho_{12}, \varphi_{12})$ and
oscillator functions in the $z_{12}$-direction
$\phi_{n_z}(z_{12})$ (for a particle of mass $\mu$), i.e.
\begin{equation}
\psi_\mathrm{rel}(\mathbf{r}_{12}) = \sum_{n,n_z} c_{n,n_z}^{(m)}
\Phi_{n,m}(\rho_{12},\varphi_{12})\,\phi_{n_z}(z_{12}).
\label{expansion_3d}
\end{equation}
The coefficients $c_{n,n_z}^{(m)}$ can be determined by
diagonalizing the Hamiltonian (\ref{relham}) in the same basis.
Evidently, in numerical analysis the basis is restricted to a
finite set $\{\Phi_{n,m}\,\phi_{n_z} |\, n = 0,\ldots,n_{\max};\,
n_z = 0,\ldots,n_z^{\max}\}$. It  must be, however, large enough
to provide a good convergence for the numerical results.

For non-interacting electrons the ground state is described by the
wave function $\psi_\mathrm{rel} = \Phi_{0,0}\,\phi_0$. For
interacting electrons, however, the ground state (in the form
(\ref{expansion_3d})) evolves from $m = 0$ to higher values of $m$
as the magnetic field strength increases. Since the quantum number
$m$ and the total spin are related by expression $S =
\frac{1}{2}[1 - (-1)^m]$, this evolution leads to the well known
singlet-triplet (S-T) transitions \cite{wag}. Note that the Zeeman
splitting (with $g^* < 0$) lowers the energy of the $M_S =1$
component of the triplet states, while leaving the singlet states
unchanged. As a consequence, the ground state is characterized by
$M_S = S$.

At the value $\omega_L^\mathrm{sph} = (\omega_z^2 -
\omega_0^2)^{1/2}$ the magnetic field gives rise to the {\it
spherical symmetry} $(\omega_z/\Omega=1)$ in the {\it
axially-symmetric} two-electron QD (with $\omega_z > \omega_0$)
\cite{hs1}.
This phenomenon was also recognized in the results for many
interacting electrons in self-assembled QDs \cite{W}.
Note that the symmetry is not approximate but {\it exact}
even for strongly interacting electrons, because the radial
electron-electron repulsion does not break the rotational
symmetry. A natural question arises how to detect such a
transition looking on the {\it ground state density distribution}
only. The related question is, if such a transition occurs, what
are the concomitant structural changes?

To this end we employ  the entanglement measure based on the
linear entropy of reduced density matrices (cf \cite{yuk})
\begin{equation}
{\cal E} = 1 - 2\,\mathrm{Tr}[{\rho_r^{(orb)}}^2]\,
\mathrm{Tr}[{\rho_r^{(spin)}}^2],
\label{enmes}
\end{equation}
where $\rho_r^{(orb)}$ and $\rho_r^{(spin)}$ are the
single-particle reduced density matrices in the orbital and spin
spaces, respectively. This measure is quite popular for the
analysis of the entanglement of two-fermion systems, in
particular, two electrons confined in the parabolic potential in
the absence of the magnetic field \cite{YPD}. Notice that
the measure (\ref{enmes}) vanishes when the global (pure) state
describing the two electrons can be expressed as one
single Slater determinant.

The trace $\mathrm{Tr}[{\rho_r^{(spin)}}^2]$ of the two-electron
spin states with a definite symmetry $\chi_{S,M_S}$ has two
values: (i) $1/2$ if $M_S = 0$ (anti-parallel spins of two
electrons); (ii) $1$ if $M_S = \pm1$ (parallel spins). The
condition $M_S = S$ yields $\mathrm{Tr}[{\rho_r^{(spin)}}^2] =
\hbox{$\frac{1}{2}$}(1 + |M_S|)=(3-(-1)^m)/4.$

The trace of the orbital part
\begin{eqnarray}
\label{oren}
\mathrm{Tr}[{\rho_r^{(orb)}}^2] \!\!&=&\!\! \int d\mathbf{r}_1\,
d\mathbf{r}_1^{\,\,\prime}\, d\mathbf{r}_2\,
d\mathbf{r}_2^{\,\,\prime}\, \psi(\mathbf{r}_1,\mathbf{r}_2)\,
\psi^*(\mathbf{r}_1^{\,\,\prime},\mathbf{r}_2) \nonumber
\\[-.5ex]
&&\qquad\quad \psi^*(\mathbf{r}_1,\mathbf{r}_2^{\,\,\prime})\,
\psi(\mathbf{r}_1^{\,\,\prime},\mathbf{r}_2^{\,\,\prime}).
\end{eqnarray}
is more involved. Indeed, in virtue of Eq.~(\ref{expansion_3d}) it
requires cumbersome calculations of eightfold sums of terms
(integrals) obtained analytically. The magnetic field dependence
of the entanglement ${\cal E}$ naturally occurs via inherent
variability of the expansion coefficients.

To characterize the Coulomb interaction strength relative to the
confinement strength we employ the so-called Wigner parameter $R_W
= (k/l_0)/\hbar\omega_0 = l_0/a^*$ (see details in \cite{lor}).
Here, $l_0
=\sqrt{\hbar/m^*\omega_0}$ is the oscillator length, and $a^* =
\hbar^2/km^*$ is the effective Bohr radius. We choose the value
$R_W = 1.5$ which corresponds for GaAs QDs to the confinement
frequency $\hbar\omega_0\approx 5.627$\,meV. 
The linear entropy
$\cal E$ is calculated using the basis with $n_{\max} = n_z^{\max}
= 4$, which gives $390625$ terms in Eq.~(\ref{oren}).

\begin{figure}
\epsfxsize 1.55in \epsffile{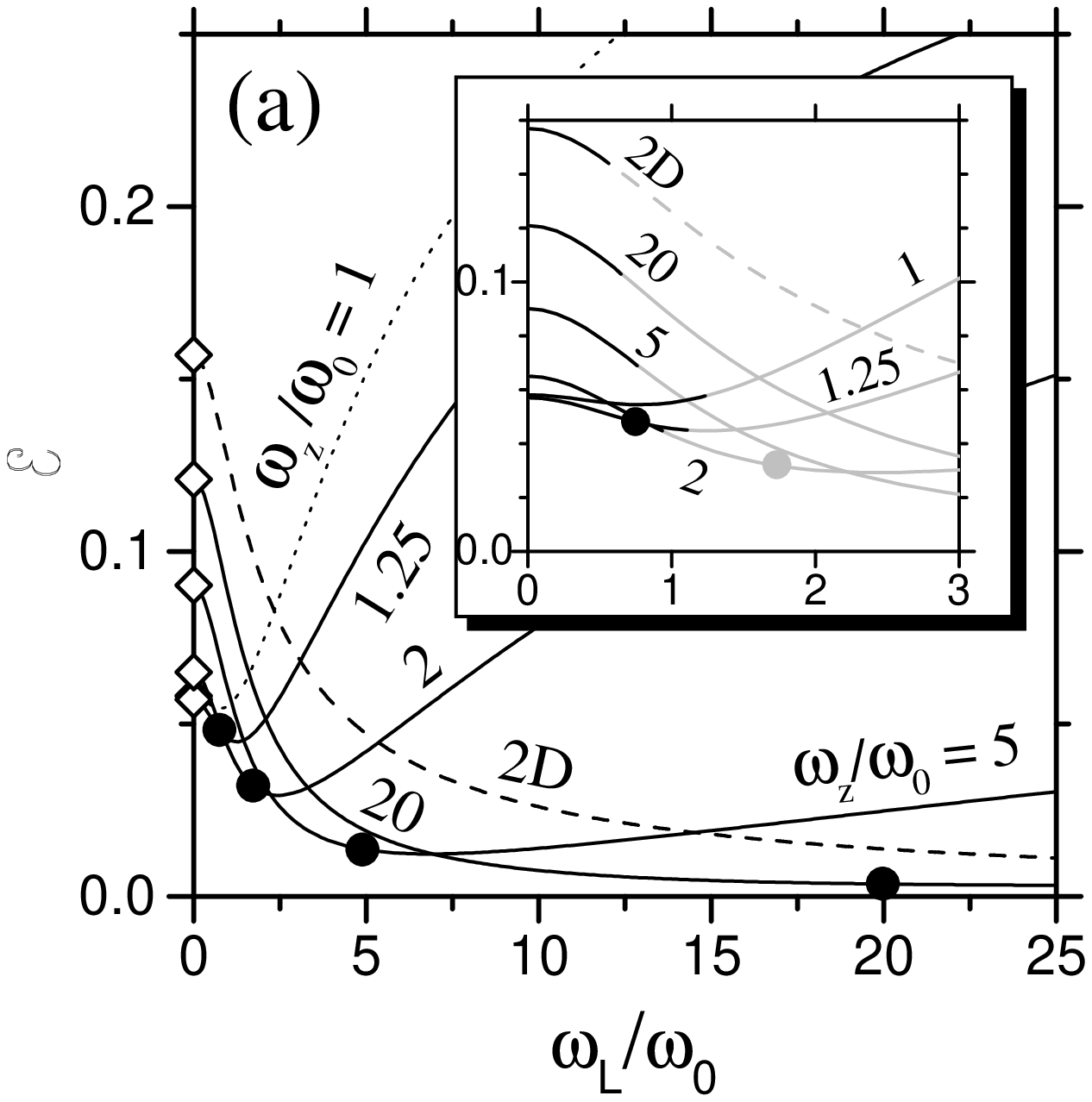}
\epsfxsize 1.55in \epsffile{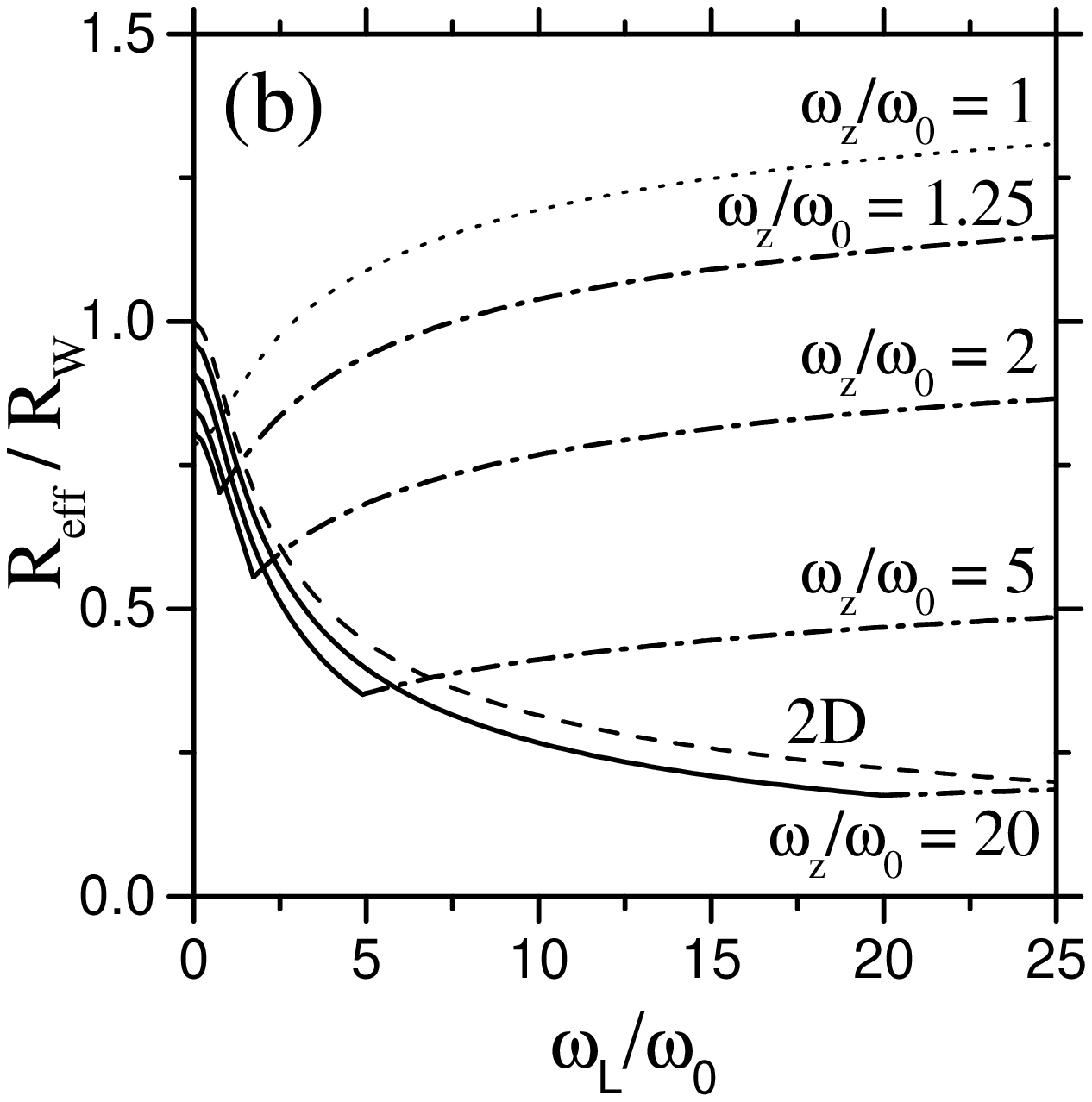}
\caption{(a) The entanglement measure $\cal E$ of the lowest state
with $m = 0$ at $R_W = 1.5$ and various ratios $\omega_z/\omega_0$
as functions of the parameter $\omega_L/\omega_0$. The initial
(black) parts of the measure (shown in the inset) correspond to
the intervals of $\omega_L/\omega_0$, when the lowest state
$m = 0$ is the ground state. Full circles denote the values of
$\omega_L/\omega_0$, when the effective 3D confinements
become spherically symmetric. (b) The relative strengths of the
Coulomb interaction $R_\mathrm{eff}^\mathrm{(2D)}/R_W$ (solid
line) and $R_\mathrm{eff}^\mathrm{(1D)}/R_W$ (dash-dotted line)
for the lowest state $m = 0$ at various ratios
$\omega_z/\omega_0$ as functions of the parameter
$\omega_L/\omega_0$.} \label{ent-oml}
\end{figure}

At zero magnetic field ($\omega_L/\omega_0 = 0$) the entanglement
of the lowest state with $m = 0$ decreases if the ratio
$\omega_z/\omega_0$ decreases from $\infty$ (2D model) to $1$
(spherically symmetric 3D model); see open symbols (diamonds) in
Fig.~\ref{ent-oml}(a). This effect could be explained by
introducing the effective charge $k_\mathrm{eff}$ \cite{ec} which
determines the effective electron-electron interaction
$V_C^\mathrm{eff} = k_\mathrm{eff}/\rho_{12}$ in the QD. In the 3D
dot the electrons can avoid each other more efficiently than in
the 2D one. Consequently, the Coulomb interaction has a smaller
effect when $\omega_0 \approx \omega_z$ (the ratio
$k_\mathrm{eff}/k \approx 0.5$) than in the anisotropic case
$\omega_0 \ll \omega_z$ ($k_\mathrm{eff}/k = 1$). Therefore, a
decreasing of the ratio $\omega_z/\omega_0$ yields an analogous
effect as the reduction of the electron-electron interaction -- a
weaker mixing of the single-particle states and, consequently, 
a lowering of the entanglement.

By increasing the magnetic field from zero to
$B_\mathrm{sph}\sim \omega_L^\mathrm{sph} = \sqrt{\omega_z^2 - \omega_0^2}$
(see full circles in Fig.~\ref{ent-oml}(a)) the
entanglement decreases. Similar to the case $B=0$, one would expect
the decrease of the effective electron-electron interaction with the evolution
of the effective confinement 
from the disk shape ($\Omega <\omega_z$)
to the spherical form ($\Omega = \omega_z$).
Further increase of the magnetic field yields the increase of the entanglement.
The effective confinement becomes again
anisotropic (now with $\Omega > \omega_z$).
Evidently, for $\omega_z/\omega_0 \to \infty$ (2D model) the minimum of $\cal E$
is shifted to infinity, i.e. in this case the entanglement
decreases monotonically with the increase of the field (dashed
line in Fig.~\ref{ent-oml}(a)).

The entanglement evolution can be explained by the influence of the
magnetic field on the effective strength of the electron-electron
interaction, which transforms the Wigner parameter $R_W$ to the
form $R_\mathrm{\Omega} = l_\Omega/a^*$. The length $l_\Omega =
\sqrt{\hbar/m^*\Omega}$ characterizes the effective lateral
confinement. 

For the quasi-2D system ($\Omega \ll \omega_z$),
the influence of magnetic field on the effective strength
$R_\Omega \Rightarrow R_\mathrm{eff}^\mathrm{(2D)} =
(k_\mathrm{eff}^\mathrm{(2D)}/l_\Omega)/\hbar\Omega$ is twofold.
Here $k_\mathrm{eff}^\mathrm{(2D)} = \langle\rho_{12} V_C \rangle$
(see Eq.~(18)-(20) in Ref.~\cite{ec}), where $V_C = k/r_{12}$ is
the full 3D Coulomb interaction. The magnetic field affects the
effective confinement  $\hbar\Omega$ as well as the effective charge. 
With the increase of the effective confinement the effective
charge $k_\mathrm{eff}^\mathrm{(2D)}/k \to 1$ and, therefore, the
effective stength decreases as $R_\mathrm{eff}^\mathrm{(2D)} \sim
1/\sqrt{\Omega}$ (see Fig.~\ref{ent-oml}(b)).

For $\Omega \gg \omega_z$ (very strong magnetic field) the
electrons are pushed laterally towards the dot's center. The
magnetic field, however, does not affect the vertical confinement.
As a consequence the electrons practically move only in the
z-direction, and the QD becomes a quasi-1D system. In this case the
effective strength is 
$R_\Omega \Rightarrow R_\mathrm{eff}^\mathrm{(1D)} =
(k_\mathrm{eff}^\mathrm{(1D)}/l_z)/\hbar\omega_z$. 
Here the
effective charge $k_\mathrm{eff}^\mathrm{(1D)} = \langle
|z_{12}|V_C\rangle$, and $l_z =
\sqrt{\hbar/m^*\omega_z}$ is the oscillator length of
the vertical confinement. For the lowest state with $m = 0$ it can be
shown that $k_\mathrm{eff}^\mathrm{(1D)}/k = (1 +
\sqrt{\omega_z/\Omega})^{-1}$. At a very strong magnetic field the
ratio $k_\mathrm{eff}^\mathrm{(1D)}/k \to 1$ which yields the
maximal value $R_\mathrm{eff}^\mathrm{(1D)} \sim
1/\sqrt{\omega_z}$.
When $\Omega = \omega_z$ the 3D system is far from both
the 2D and the 1D limits. As a result,
$R_\mathrm{eff}^\mathrm{(2D)}$ and $R_\mathrm{eff}^\mathrm{(1D)}$
do not match smoothly (see Fig.~\ref{ent-oml}(b)). However, it is
clear that the effective strength reaches the minimum around this
point, i.e. when the transition from the lateral to the vertical
localization of two electrons takes place.

\begin{figure}
\epsfxsize 3.25in \epsffile{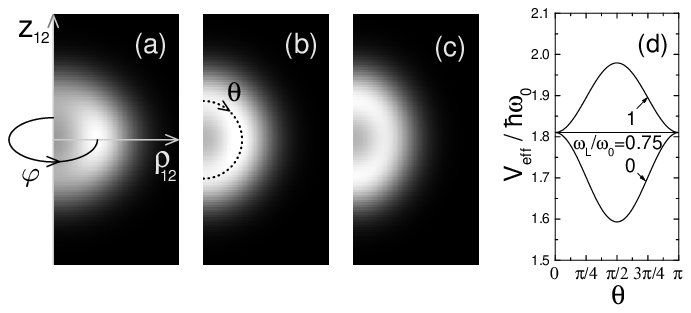}
\caption{The ground state probability density (the relative motion
part, $m = 0$) for the near-spherical ($\omega_z/\omega_0 = 1.25$)
two-electron QD with $R_W = 1.5$ ($\hbar\omega_0 = 5.627$\,meV)
and $g^* = -0.44$ at different values of the magnetic field: (a)
$\omega_L/\omega_0 = 0$, (b) $\omega_L/\omega_0 = 0.75$, (c)
$\omega_L/\omega_0 = 1$. For the spherically symmetric case (b)
the maximum of the probability density spreads uniformly over the
spherical shell of a radius $r_0$. The panel (d) displays the
potential $V_\mathrm{eff}(\rho_{12},z_{12})$ on this sphere
($\rho_{12} = r_0\sin\theta$, $z_{12} = r_0\cos\theta$) for the
cases (a-c).} \label{fig2}
\end{figure}

In order to get a deeper insight into this transition we examine
the probability density $\vert\psi({\bf r}_{12})\vert^2$ for the
ground state, when $m = 0$. Such a ground state can be realized  
for a near spherical QD with $\omega_z/\omega_0 = 1.25$. 
For this ratio and $R_W=1.5$
the first S-T transition occurs at $\omega_L/\omega_0 \approx 1.11$
(see the inset in Fig.~\ref{ent-oml}(a)). The
spherical symmetry ($\Omega = \omega_z$) sets up
at $\omega_L^{\rm sph}/\omega_0 = 0.75$, i.e., in the ground state.
For the magnetic field strengths  $\omega_L < \omega_L^{\rm sph}$
($\Omega < \omega_z$) the density maximum forms a ring in
the $(x_{12},y_{12})$-plane (a consequence of the axial symmetry).
Fig.~\ref{fig2}(a) shows the cut of this density with the
$(\rho_{12},z_{12})$-plane at arbitrary azimuthal angle $\varphi$.
For $\omega_L = \omega_L^{\rm sph}$ the maximum of
the probability density forms a spherical shell (visible if we
rotate Fig.~\ref{fig2}(b) around $z_{12}$-axis).
For $\omega_L > \omega_L^{\rm sph}$ ($\Omega > \omega_z$)
two separate density maxima start to grow, located symmetrically
in the $z_{12}$-axis (see Fig.~\ref{fig2}(c)).  In
contrast to the corresponding behaviour of the entanglement, a fuzzy transition 
manifests itself in the probability density   
for the chosen dot's parameters. In fact, the entanglement 
evolution guides us to trace a geometrical crossover from the lateral
to the vertical localization of the electrons.

The probability density evolution due to the magnetic field shown in
Figs.~\ref{fig2}(a-c) can be elucidated by means of the analysis of the
effective potential $V_\mathrm{eff} = \frac{1}{2}\,\mu\,(\Omega^2
\rho_{12}^2 + \omega_z^2 z_{12}^2) + k/r_{12} +
\hbar^2m^2/(2\mu\rho_{12}^2)$. Namely, the maxima of the
probability density for the ground state are directly related to
the minima of $V_\mathrm{eff}$. For $\omega_L < \omega_L^{\rm
sph}$ the potential surface has the minimum at $\rho_{12} =
\rho_0$, $z_{12} = 0$, where $\rho_0 = (k/\mu\Omega^2)^{1/3}$ if
$m = 0$. By increasing the magnetic field to values
$\omega_L > \omega_L^{\rm sph}$ this minimum transforms to the
saddle point in the $(\rho_{12},z_{12})$-plane, but two new minima
divided by this saddle (potential barrier) appear. For $m = 0$
these minima are located at $z_{12} =
\pm z_0$, where $z_0 = (k/\mu\omega_z^2)^{1/3}$.

For weakly anisotropic (near spherical) systems it is convenient
to use the spherical coordinates $(r_{12},\theta,\varphi)$, where
the polar angle is $\theta \equiv \arctan(\rho_{12}/z_{12})$. In
these coordinates the positions of the minima are $r_{12} =
\rho_0$, $\theta = \pi/2$ for $\Omega < \omega_z$ and $r_{12} =
z_0$, $\theta = 0, \pi$ for $\Omega > \omega_z$; see the cases
$\omega_L/\omega_0 = 0$ and $1$, respectively, in
Fig.~\ref{fig2}(d). Note, that due to the axial symmetry the
azimuthal angle $\varphi$ is arbitrary. The maximum at $\theta =
\pi/2$ for $\Omega > \omega_z$ (the case $\omega_L/\omega_0 = 1$
in Fig.~\ref{fig2}(d)) corresponds to the saddle point at $z_{12}
= 0$. If we consider small oscillations around a minimum, the
effective potential can be written in the form $V_\mathrm{eff}
\approx V_0 + \frac{1}{2}\,\omega_1^2\,q_1^2 +
\frac{1}{2}\,\omega_2^2\,q_2^2$ (the expansion up to the quadratic
terms), where $q_1 = \Delta r$, $q_2 = r_0 \Delta\theta$ are the
normal coordinates. Here $r_0 = \rho_0$ when $\Omega < \omega_z$,
whereas $r_0 = z_0$ when $\Omega > \omega_z$. The corresponding
normal frequencies (if $m = 0$) are: (i) $\omega_1 =
\sqrt{3}\,\Omega$, $\omega_2 = (\omega_z^2 - \Omega^2)^{1/2}$ for
$\Omega < \omega_z$; and (ii) $\omega_1 = \sqrt{3}\,\omega_z$,
$\omega_2 = 2\,(\Omega^2 - \omega_z^2)^{1/2}$ for $\Omega >
\omega_z$. For the spherically symmetric case ($\Omega =
\omega_z$) one has $\omega_2 = 0$ and the minima of
$V_\mathrm{eff}$ degenerate to the sphere of radius $r_0 = \rho_0
= z_0$. In other words, the potential $V_\mathrm{eff}$ becomes
independent on the angle $\theta$; see the case $\omega_L/\omega_0
= 0.75$ in Fig.~\ref{fig2}(d). As a consequence, the wave function
becomes spherically symmetric (Fig.~\ref{fig2}(b)).
The quantum oscillations evolve in a way similar to those     
of quantum phase transitions studied for model systems \cite{Sad}. 

In conclusion, we found that the entanglement of the lowest state
with $m=0$ in  axially symmetric two-electron QDs, being first a
decreasing function of the magnetic field, starts to increase
after the transition point $\omega_L^\mathrm{sph} =
\sqrt{\omega_z^2 - \omega_0^2}$ with the increase of the magnetic
field. This behaviour is understood as the transition from the
lateral to the vertical localization of  two electrons. 
It is especially noteworthy that the transition point
associated with the onset of the spherical symmetry is robust at
any strength of the Coulomb interaction at the fixed ratio of the
quantum confinement  $\omega_z/\omega_0$. Varying the magnetic
field around the transition point, one can control the
increase/decrease of the entanglement in QDs.
We paid a special attention to the case when the shape transition
occurs in the dot's ground state. This can happen for weakly
anisotropic QDs (see the inset in Fig.~\ref{ent-oml}(a)). Note
that for a typical ratio $\omega_z \gg \omega_0$ considered in
literature the lowest $m = 0$ state becomes an excited state
before the shape transition takes place. Therefore, it would be
difficult to recognize such type of transitions in previously
considered cases. We speculate that the transition may be
observed in optical experiments, where the appearance of the wave
function in the vertical direction may affect the
photoluminescence of the QD. However, this question requires a
dedicated study and is beyond the scope of the present paper.

This work is partly supported by RFBR Grant No.11-02-00086 (Russia), Project
171020 of Ministry of Education and Science of Serbia;
by the CAIB and FEDER (Spain).

\end{document}